\documentclass[aps, pra, showpacs, twocolumn, amsfonts, amsmath, amssymb, floatfix]{revtex4}

\usepackage{graphicx}

\begin{document}

\title{Fast optimal transition between two equilibrium states}

\author{Jean-Fran\c{c}ois Schaff}
\affiliation{Universit\'e de Nice-Sophia Antipolis, Institut Non Lin\'eaire de Nice, CNRS, 1361 route des Lucioles, F-06560 Valbonne, France}

\author{Xiao-Li Song}
\affiliation{Universit\'e de Nice-Sophia Antipolis, Institut Non Lin\'eaire de Nice, CNRS, 1361 route des Lucioles, F-06560 Valbonne, France}

\author{Patrizia Vignolo}
\affiliation{Universit\'e de Nice-Sophia Antipolis, Institut Non Lin\'eaire de Nice, CNRS, 1361 route des Lucioles, F-06560 Valbonne, France}

\author{Guillaume Labeyrie}
\email{guillaume.labeyrie@inln.cnrs.fr}
\affiliation{Universit\'e de Nice-Sophia Antipolis, Institut Non Lin\'eaire de Nice, CNRS, 1361 route des Lucioles, F-06560 Valbonne, France}


\begin{abstract}
We demonstrate a technique based on invariants of motion for a time-dependent Hamiltonian, allowing a fast transition to a final state identical in theory to that obtained through a perfectly adiabatic transformation. This method is experimentally applied to the fast decompression of an ultracold cloud of Rubidium 87 atoms held in a harmonic magnetic trap, in the presence of gravity. We are able to decompress the trap by a factor of 15 within 35 ms with a strong suppression of the sloshing and breathing modes induced by the large vertical displacement and curvature reduction of the trap. When compared to a standard linear decompression, we achieve a gain of a factor of 37 on the transition time.
\end{abstract}

\maketitle
\section{Introduction}

The controlled manipulation of quantum states is central to many areas of physics such as quantum information processing~\cite{Calarco2004,Dechiara2008}, design of pulses for nuclear magnetic resonance imaging~\cite{Garroway1974,Sutherland1978,Hoult1979}, atomic gas cooling~\cite{Leanhardt2003} and transport~\cite{Esslinger2001}, or ion manipulation~\cite{Schulz2006}.
The paradigm of adiabatic transformations, in which the Hamiltonian parameters are changed infinitesimally slowly with time~\cite{Kato1950}, is often used to drive a system from a given quantum state to another. However, the urge to shorten the duration of the experiments has driven the search for fast optimal non-adiabatic strategies~\cite{Bulatov1998,Schulz2006,DGO2008,Salamon2009,Chen2010}, with a minimal amount of extra energy supplied to the system.

In the particular field of cold atoms, time-dependent potentials are becoming increasingly used. Examples include the transport of cold atomic samples over various distances~\cite{Esslinger2001,Cornell2003,DGO2008}, or the production of very low temperatures using trap decompression~\cite{Leanhardt2003}. To minimize the energy imparted to the atoms, most of these experiments were performed in the adiabatic regime where the process duration was much longer than the oscillation period in the potential, yielding times in the few seconds range or longer. Achieving a faster transfer with a limited heating motivated experimentalists to employ various non-adiabatic procedures~\cite{DGO2008, Wang2010}. Recently, a method based on invariants of motion was proposed for the decompression of harmonic traps which was argued to give access to shorter times than ``bang-bang'' control~\cite{Salamon2009}, provided that negative curvatures could be transiently applied~\cite{Chen2010}. 

We present in this article an experimental demonstration of shortcuts to adiabaticity based on this method, which we employ to decompress a cloud of magnetically trapped $^{87}$Rb atoms. Because of gravity, the position of the trap center shifts vertically, which induces sloshing modes of the trapped atoms. At the same time, a breathing mode is excited by the reduction of the trap frequency.
We thus generalize the approach of Ref.~\cite{Chen2010} to the case of a time-dependent harmonic plus constant linear potential to account for gravity. We derive a trap frequency trajectory $\omega_z(t)$ which yields a final state identical to that obtained through a purely adiabatic transformation (hence the ``optimal'' transition), but in a much shorter time. We experimentally implement this trajectory to perform a vertical trap decompression by a factor of 15 within 35 ms (corresponding to roughly half the decompressed trap oscillation period), with a strong suppression of the cloud's center of mass and size oscillations.

\section{Optimal trajectory determination}
We start our theoretical approach by considering a time-dependent harmonic oscillator in the presence of gravity
\begin{equation}
H(t) = \frac{p^2}{2m} + \frac{1}{2}m\omega_z^2(t)z^2 + mgz,
\end{equation}
with initial and final angular frequencies $\omega_z(0)=\omega_{0z}$ and $\omega_z(t_f)=\omega_{fz}$, respectively. The objective is to engineer a trajectory $\omega_z(t)$ between these two values so that if we start with an initial state at equilibrium at temperature $T_0$, this state is mapped to a final equilibrium state at temperature $T_f=T_0/\gamma^2$, with $\gamma^2=\omega_{0z}/\omega_{fz}$ \cite{Chen2010}.
Our solution is based on invariants of motion of the form \cite{Leach,Schaff2010}
\begin{equation}
I(t)=\dfrac{\Pi^2}{2m}+\dfrac{1}{2}m\omega_{0z}^2 Q^2,
\label{invariant}
\end{equation}
where $Q=z/b+ga/\omega_{0z}^2$ and $\Pi=b p-m\dot b  z+m b^2g\dot a/\omega_{0z}^2$ play the role of canonical variables.
For Eq.~(\ref{invariant}) to be an invariant, the dimensionless functions $b$ and $a$, respectively linked to the size $\sigma_z$ and center-of-mass position $z_{cm}$ of the cloud through $\sigma_z(t)=b(t)\sigma_z(0)$ and $z_{cm} (t)= - a(t)b(t)g/\omega_{0z}^2$, must be solutions of
\begin{gather}
\label{eq:b}
d^2 b/d t^2 + b(t)\omega_z^2(t) = \omega_{0z}^2/b(t)^3, \\
\label{eq:a}
d^2 a/d \tau^2 + a(\tau) = b(\tau)^3,
\end{gather}
where $\tau(t) = \omega_{0z}\int_0^t dt'/b^2$.
%The dynamical states coincide with the corresponding eigenstates of the initial and final Hamiltonians
The solutions of the time-dependent Schr\"odinger equation coincide with the stationary states of the initial and final Hamiltonians $H(t=0)$ and $H(t_f)$ if $I(t=\{0,t_f\})\propto H(t=\{0,t_f\})$ \cite{Lewis1969}. Thus we set $\dot a(0)=\dot a(t_f)=\dot b(0)=\dot b(t_f)=0$ and $a(0)=1$, $a(t_f)=\gamma^3$, $b(0)=1$, $b(t_f)=\gamma$. These latter terms imply that $\ddot b(0)=\ddot b(t_f)=0$ must hold as well, giving ten independent boundary conditions (BC).
Our procedure to engineer $\omega_z(t)$ is the following: (i) we use a polynomial ansatz for $a(\tau)$ of the form $a(\tau)=\sum_{j=0}^{j\ge9}\alpha_j(\tau/\tau_f)^j$, for which ten coefficients are fixed by the BC and the other can be arbitrarily chosen; (ii) we evaluate $b^3(\tau)$ and thus $b[\tau(t)]$; and (iii) using Eq.~\eqref{eq:b} we obtain the function $\omega_z(t)$. Quite non-intuitively, the obtained solution is valid for \emph{any magnitude} of the linear term in the time-dependent Hamiltonian (as long as this linear term is not time-dependent itself). In the particular case of $g = 0$ (no constant force term), however, a lower-order polynomial ansatz (fifth order) is sufficient~\cite{Chen2010}.

\begin{figure}
\begin{center}
\resizebox{0.8\columnwidth}{!}{\includegraphics{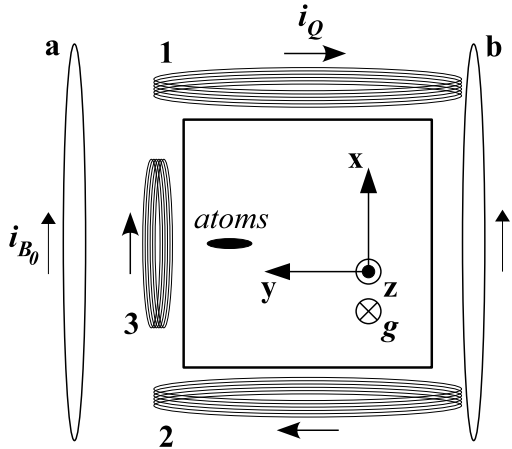}} 
\caption{Trapping geometry (figure in the horizontal plane). Ultracold $^{87}$Rb atoms are trapped in an Ioffe-Pritchard-type magnetic trap created by current $i_Q$ running through the three QUIC coils \textbf{1}, \textbf{2}, and \textbf{3}. An additional pair of coils (\textbf{a} and \textbf{b}) produces an homogeneous field along $y$, which allows an independent tuning of the trap minimum field $B_0$ via the current $i_{B_0}$.}
\label{fig1}
\end{center}
\end{figure}

\section{Experimental procedure}
To experimentally investigate shortcuts to adiabaticity, we employ a sample of ultracold $^{87}$Rb atoms held in a magnetic Ioffe-Pritchard trap. This popular type of trap is harmonic (for cold enough atoms) and anisotropic, with a typical ratio of 10 between the oscillation frequencies in the radial dimensions $\omega_{x,z}$ and the axial one $\omega_y$ (see Fig.~\ref{fig1}) yielding the well-known cigar-shaped aspect of the trapped cloud. For shallow traps, gravity significantly affects the potential in the vertical dimension, yielding a \emph{displacement} of the trap minimum $-g/\omega_z^2$ compared to a tight trap. Our magnetic trap is of the quadrupole-Ioffe-configuration type (QUIC trap) introduced in Ref.~\cite{Hansch1998}, the three-coils setup sketched in Fig.~\ref{fig1}. For sufficiently cold atoms ($k_B T \ll \mu B_0$), the magnetic potential is harmonic of the form~\cite{Foot2002} 
\begin{equation}
\mu B = \mu \left[B_0 + \frac{1}{2} \left(\frac{B'^2}{B_0}-\frac{B''}{2}\right) \left(x^2 + z^2\right) + \frac{1}{2} B'' y^2\right],
\label{IP}
\end{equation}
where $\mu/h \approx 1.4$ MHz/G for our atoms in $\left|F=2, m_F=+2\right\rangle$. $B'$ is the radial gradient of the magnetic field while $B''$ represents its curvature along \textit{y}. $B_0$ is the minimum of the magnetic field at the trap center, which can be adjusted using two independent parameters: the current $i_Q$ running in the three QUIC coils, or the current $i_{B_0}$ in a pair of compensation coils providing a uniform field along $y$ (see Fig.~\ref{fig1}). Since $B'' \ll B'^2/B_0$, the radial and axial angular frequencies are given by
\begin{gather}
\label{omega_r}
\omega_{x,z} \approx \sqrt{\frac{\mu}{m}} \frac{B'(i_Q)}{\sqrt{B_0(i_Q, i_{B_0})}} \, ,\\
\label{omega_a}
\omega_y = \sqrt{\frac{\mu}{m}}\sqrt{B''(i_Q)} \, .
\end{gather}
These expressions show that we can, to some extent, manipulate independently the radial and axial frequencies using $i_Q$ and $i_{B_0}$.

Our initial sample is a small ($N=10^5$ atoms) and cold ($T_0 = 1.63$ $\mu$K) atomic cloud. The low temperature guarantees that the potential seen by the atoms remains harmonic even for large decompression factors. The small number of atoms is chosen to reduce the density and thus the elastic collision rate, responsible for the energy transfer between dimensions and thermalization. In the compressed trap with previously mentioned parameters, the typical time between two elastic collisions is $\approx 28$ ms, quite larger than the radial oscillation period of $4$ ms.

Prior to implementing a decompression sequence, we need to characterize the initial and final states. To this end, the position and size of the atomic cloud in three dimensions (3D) are measured using absorption imaging along two orthogonal directions. The trap frequencies are measured by slightly offsetting the trap center using compensation coils, then abruptly releasing it and measuring the cloud's center-of-mass motion as a function of time. For our fully compressed trap ($i_Q=27$ A, $i_{B_0}=0$), we obtain $\nu_{0x} = \omega_{0x} / 2 \pi = 228.1$ Hz, $\nu_{0y} = 22.2$ Hz and $\nu_{0z} = 235.8$ Hz. To measure the parameters of the final, decompressed state, we perform an ``adiabatic-like'' (i.e., slow $t_f = 6$ s) decompression using linear ramps for the currents $i_Q$ and $i_{B_0}$. In the following, we will refer to such ramps as ``linear decompressions,'' although the resulting $\omega_z(t)$ is not strictly linear [Eqs.~(\ref{omega_r}) and~(\ref{omega_a})]. The results presented in this article are obtained with a vertical decompression factor $\nu_{0z}/\nu_{fz} = 15$, yielding final frequencies $\nu_{fx} = 18.1$ Hz, $\nu_{fy} = 7.1$ Hz, and $\nu_{fz} = 15.7$ Hz for the decompressed trap. In practice, this is achieved by decreasing $i_Q$ from $27$ to $3.6$ A and increasing $i_{B_0}$ from $0$ to $3$ A. Since $\nu_y$ is not affected by the increase of $i_{B_0}$ [see Eq.~(\ref{omega_a})], the decompressed trap is much more isotropic ($\nu_{f\{x,z\}} / \nu_{fy} \approx 2$) than the compressed one ($\nu_{0\{x,z\}} / \nu_{0y} \approx 12$).

\begin{figure}
\begin{center}
\resizebox{0.95\columnwidth}{!}{\includegraphics{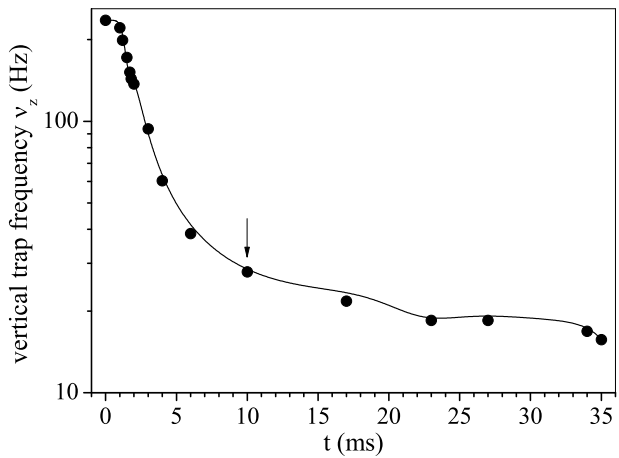}} 
\caption{Optimal trap frequency trajectory for a 35 ms vertical decompression. We plot (line) $\nu_z(t)$ for a 35 ms vertical decompression from $\nu_{0z} = 235.8$ Hz to $\nu_{fz} = 15.7$ Hz, obtained with the invariant method (see text). The symbols correspond to measured values of the vertical trap frequency during the decompression process.}
\label{fig2}
\end{center}
\end{figure}

We illustrate the efficiency of our shortcut method by realizing a fast ($t_f = 35$ ms) trap decompression optimized for the vertical dimension $z$, where gravity strongly affects the cloud's motion. The employed solution $\nu_z (t)$ is shown in Fig.~\ref{fig2} (line, note the vertical log scale). 
%It was obtained using the general procedure exposed before, with the addition of one degree of freedom to achieve a reasonably smooth field shape. %Hence, an 11th-order polynomial was employed.
Because of the finite time response of the trap electronic circuit, the measured trap field profile is different from the computed one. We thus monitored $\nu_z$ by interrupting the sequence at different times, and adjusted the compensation field to obtain a measured $\nu_z (t)$ (symbols in Fig.~\ref{fig2}) close to the theoretical one (deviation $<5\%$). The uncertainty on the experimental values is $\pm 2\%$.

\begin{figure}
\begin{center}
\resizebox{1.05\columnwidth}{!}{\includegraphics{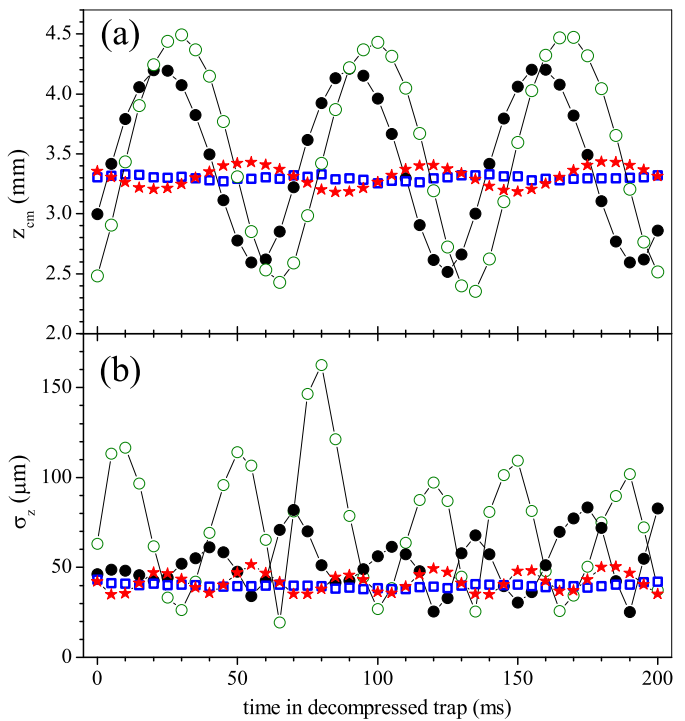}} 
\caption{(Color online) Vertical trap decompression: comparison between different schemes. We report in (a) and (b), respectively, the cloud's vertical center-of-mass position $z_{cm}$ and size $\sigma_z$ versus time after decompression, for four different sequences. Open circles (green): abrupt decompression; solid circles (black): linear decompression in 35~ms; stars (red): shortcut decompression in 35~ms; squares (blue): linear decompression in 6~s.}
\label{fig3}
\end{center}
\end{figure}

\section{Results}
Figure~\ref{fig3} shows the result of the shortcut decompression using the trajectory of Fig.~\ref{fig2}. We plot in Fig.~\ref{fig3}(a) the time evolution of the cloud's center-of-mass position $z_{cm}$ once the decompression sequence is completed, and in Fig.~\ref{fig3}(b) that of the cloud's size $\sigma_z$. These data correspond to averages over three successive images, taken after a 6~ms time~of~flight. The open circles correspond to an abrupt jump from $\nu_{0z}$ to $\nu_{fz}$ (in practice, the effective decompression time is $\approx 0.1$ ms). The solid circles are obtained with a (non-optimal) linear decompression of duration $t_f = 35$ ms, the stars with the shortcut trajectory, and the squares with a quasi-adiabatic linear decompression in 6~s. In every instance, we observe in Fig.~\ref{fig3}(a) the expected sinusoidal oscillations of $z_{cm}$ at the decompressed trap frequency $\nu_{fz} = 15.7$ Hz (dipole mode) and of amplitude $\Delta z_{cm}$. As can be seen, the shortcut decompression yields a strong reduction of $\Delta z_{cm}$ when compared to the abrupt and 35~ms linear decompressions, by a factor 9 and 7.2, respectively. However, the residual center-of-mass oscillations after the shortcut sequence are still sizable, a factor of 5 larger than that observed for the 6-s-long linear decompression. We attribute these residual oscillations to imperfections of the experiments which are discussed at the end of the paper. From the amplitude $\Delta z_{cm}$ we can infer the excess energy communicated to the cloud in the form of the dipole excitation $E_{dip} = 1/2 m \omega_{fz}^2 {\Delta z_{cm}}^2$. We also observe in Fig.~\ref{fig3}(b) oscillations of the cloud's size $\sigma_z$ at twice the frequency of the decompressed trap (breathing mode). The theory predicts such non-sinusoidal periodic oscillations, whose expression can be derived analytically ~\cite{Minguzzi2005}. 
In the experiment, the measurement of $\sigma_z$ is less accurate than that of $z_{cm}$, because of the limited spatial resolution and noise ($\sigma_z \approx 40 \mu$m), and we cannot fit the measured oscillations to the model in every instance. We thus quantify the amplitude of the breathing mode by using the standard deviation $\Delta \sigma_z$ of $\sigma_z (t)$ after decompression. We observe a reduction of $\Delta \sigma_z$ when we use the shortcut trajectory, by a factor $7$ and $3$ when compared to the abrupt and linear decompressions respectively. The residual $\Delta \sigma_z$ is again a factor of 5 above that of the 6-s-long linear decompression. The excess energy stored in the breathing mode is $E_{breath} \approx 2 m \omega_{fz}^2 {\Delta \sigma_z}^2$. Note that we overestimate $\Delta \sigma_z$ (and thus $E_{breath}$) because of our 6~ms time of flight. The total excess energy imparted to the system during the decompression is then $E_{exc} = E_{dip} + E_{breath}$. Quite obviously from the vertical scales in Fig.~\ref{fig3}, we always have $E_{dip} \gg E_{breath}$. Using the previous expression, we find excess energies of 54, 35, 0.7, and 0.02~$\mu$K for the abrupt, 35~ms linear, 35~ms shortcut, and 6~s linear decompressions, respectively. For the latter, we measured a final temperature $T_f = 0.13~\mu$K. Since the initial temperature is $T_0 = 1.63~\mu$K, the cooling factor is 12.5, quite close to the expected $\nu_{0z}/\nu_{fz} = 15$ value for a purely adiabatic transition.

\begin{figure}
\begin{center}
\resizebox{1.0\columnwidth}{!}{\includegraphics{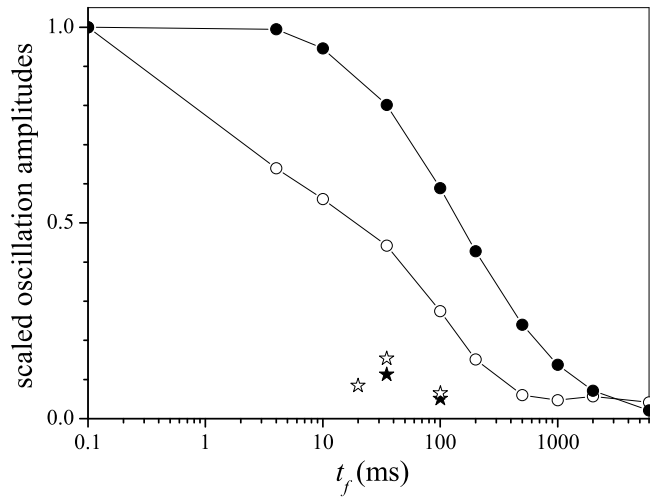}} 
\caption{Summary of faster-than-adiabatic decompression results. We plot the amplitudes of center-of-mass (filled circles) and cloud's size (open circles) oscillations along the vertical direction after linear decompressions of various durations $t_f$. All amplitudes are scaled to that of the abrupt decompression ($t_f = 0.1$~ms). The stars correspond to our shortcut decompression experiments in 20, 35 and 100~ms (filled symbols: center of mass, open symbols: size).}
\label{fig4}
\end{center}
\end{figure}

To provide the reader with a better feeling of the time scales involved in the trap decompression, we compare in Fig.~\ref{fig4} our shortcut results with those of linear decompressions with various durations (full circles for $\Delta z_{cm}$, open circles for $\Delta \sigma_z$). All the amplitudes in this figure are normalized to those corresponding to an abrupt decompression ($t_f = 0.1$ ms). 
The stars correspond to three shortcut experiments. Two experiments were performed along the vertical: the 35~ms one depicted on Figs.~\ref{fig2} and~\ref{fig3}, and a 100-ms-long one. Another, 20-ms-long shortcut decompression was also performed along $x$ (no gravity), using the fifth-order polynomial ansatz of Ref.~\cite{Chen2010}. The solid stars correspond to oscillation amplitudes of $z_{cm}$ while the open stars stand for $\Delta \sigma_z$. The linear decompression data allow us to estimate a \textit{quantitative} criterion for adiabaticity instead of the usual qualitative criterion $t_f \gg 1/\omega$. For instance, we can set as a criterion that the excess energy should be of the order or smaller than the thermal energy associated with $T_f = T_0/\gamma^2$: $E_{exc} \simeq 1/2 m \omega_{fz}^2 {\Delta z_{cm}}^2 \leq k_B T_f$. This condition yields $t_f \geq 3.3$ s for linear ramps. Since our shortcut sequence in $100$ ms also satisfies this condition it can be considered adiabatic using this criterion, with a reduction of the necessary decompression time by a factor of 33. The 35 ms shortcut decompression does not meet the criterion above but still realizes a gain on the transition time of a factor of 37 when compared to a linear ramp. The 20-ms-long decompression along $x$ excites only the breathing mode (no trap displacement), whose residual amplitude is one order of magnitude lower than for the abrupt transition, and a factor of 2 above that of the 6 s linear ramp.            

As stressed in the theoretical part of this article, only the \emph{final state} is identical to that obtained through an adiabatic sequence. Indeed, we performed an experiment where we interrupted the frequency trajectory of Fig.~\ref{fig2} after 10 ms (as pointed out by the arrow in the figure). Despite the fact that 94$\%$ of the frequency difference $\nu_{0z}-\nu_{fz}$ has been covered at $t = 10$ ms, we observe large center-of-mass and cloud size oscillations, respectively a factor of 7 and 4 larger than those observed when the entire 35~ms sequence is completed. Thus, the last 25~ms of the frequency trajectory in Fig.~\ref{fig2} are of paramount importance for reaching the optimal final state.

We now discuss experimental imperfections which might be responsible for the residual oscillations observed in Fig.~\ref{fig3}. The first possible cause is a mismatch between the theoretical frequency trajectory and the experimental one. As shown in Fig.~\ref{fig2}, we did our best to maintain this mismatch below 5$\%$ for selected time values of the trajectory, but we cannot guarantee that this holds for the whole sequence. In particular, as discussed previously, the last part of the trajectory where the frequencies are small and thus the relative measurement error large is potentially more critical.
Probably most importantly, our trap can be considered harmonic only for small atomic displacements from the trap center. During the shortcut decompression, the trap center shifts vertically by $\approx$ 1~mm and the atoms follow a complex dynamics that brings them quite far from the trap center ($\approx 300 \mu$m). Deviations from harmonicity may thus play an important role in our experiment~\cite{DGO2008}, limiting the performances of our shortcut decompression.

\section{Conclusion}
In conclusion, we presented in this article the first experimental realization of the faster-than-adiabatic displacement and cooling of an ensemble of magnetically trapped ultracold atoms using an optimal decompression sequence based on invariants of motion. Using this formalism, we derived optimal trap frequency trajectories in the case of a time-dependent harmonic potential plus a time-independent linear term accounting for gravity. Our solution also applies to the simpler case of a purely harmonic potential such as that treated in Ref.~\cite{Chen2010}. We demonstrated the validity of our scheme by applying a fast (35~ms) 15-fold frequency decompression to the trap in the vertical dimension, yielding a residual center-of-mass oscillation of the cloud equivalent to that of 1.3-s-long linear decompression (a reduction by a factor of 37). As a future prospect, one could apply this technique to more isotropic traps (such as crossed dipole traps) to obtain a faster and efficient cooling in 3D and produce very low temperatures.
Optimal trajectories could also be searched for in other situations such as the moving quadrupole magnetic traps often used to transport cold atoms~\cite{Esslinger2001}. This method can also be readily applied to a Tonks gas~\cite{Minguzzi2005}, and to Bose-Einstein condensates with some restrictions on the dimensionality due to the scaling of the interaction term~\cite{Kagan1996,DGO2009}. More generally, these optimal faster-than-adiabatic schemes could be adapted to many areas of physics where time-dependent Hamiltonians are employed.

\section{Acknowledgments}
We thank J.-C. Bery, J.-C. Bernard and A. Dusaucy for their assistance in setting up the BEC experiment. This work was supported by CNRS and Universit\'e de Nice-Sophia Antipolis. We also acknowledge financial support from R\'egion PACA and F\'ed\'eration Wolfgang Doeblin.


\begin{thebibliography}{10}
\bibitem{Calarco2004} T. Calarco, U. Dorner, P.S. Julienne, C.J. Williams, and P. Zoller, Phys. Rev. A {\bf 70}, 012306 (2004).
\bibitem{Dechiara2008} G. De Chiara, T. Calarco, M. Anderlini, S. Montangero, P.J. Lee, B.L. Brown, W.D. Phillips, and J.V. Porto, Phys. Rev A {\bf 77}, 052333 (2008).
\bibitem{Garroway1974} A. Garroway, P. Grannell, and P. Mansfield, J. Phys. C {\bf 7}, L457 (1974).
\bibitem{Sutherland1978} R. Sutherland and J. Hutchison, J. Phys. E {\bf 11}, 79 (1978).
\bibitem{Hoult1979} D. Hoult, J. Magn. Reson. {\bf 35}, 69 (1979).
\bibitem{Leanhardt2003} A.E. Leanhardt, T.A. Pasquini, M. Saba, A.Schirotzek, Y. Shin, D. Kielpinski, D.E. Pritchard, and W. Ketterle, Science {\bf 301}, 1513 (2003).
\bibitem{Esslinger2001} M. Greiner, I. Bloch, T. W. Hansch and T. Esslinger, Phys. Rev. A \textbf{63}, 031401 (2001).
\bibitem{Schulz2006} S. Schulz, U. Poschinger, K. Singer, F. Schmidt-Kaler, Fortschr. Phys. {\bf 54} 648 (2006).
\bibitem{Kato1950} T. Kato, J. Phys. Soc. Jpn. {\bf 5}, 435 (1950).
\bibitem{Bulatov1998} A. Bulatov, B. Vugmeister, A. Burin, and H. Rabitz, Phys. Rev. A {\bf 58}, 1346 (1998).
\bibitem{DGO2008} A. Couvert, T. Kawalec, G. Reinaudi, and D. Gu\'ery-Odelin, Euro. Phys. Lett. \textbf{83}, 13001 (2008).
\bibitem{Salamon2009} P. Salamon, K. Heinz Hoffmann, Y. Rezek and R. Kosloff, Phys. Chem. Chem. Phys. {\bf 11}, 1027 (2009).
\bibitem{Chen2010} X. Chen, A. Ruschhaupt, S. Schmidt, A. del Campo, D. Gu\'ery-Odelin, and J.G. Muga, Phys. Rev. Lett. \textbf{104}, 063002 (2010).
\bibitem{Cornell2003} H. Lewandowski, D. Harber, D. Whitaker and E. Cornell, J. Low Temp. Phys. \textbf{132}, 309 (2003).
\bibitem{Wang2010} D. Chen, H. Zhang, X. Xu, T. Li, and Y. Wang, Appl. Phys. Lett. \textbf{96}, 134103 (2010).
\bibitem{Leach} P.G.L. Leach, J. Math. Phys. {\bf 18}, 1608 (1977); P.G.L. Leach, J. Math. Phys. {\bf 18}, 1902 (1977); P.G.L. Leach, Am. J. Phys. {\bf 46}, 1247 (1978).
\bibitem{Schaff2010} J.-F. Schaff {\it et al.}, (unpublished).
\bibitem{Lewis1969} H.R. Lewis, Jr. and W.B. Riesenfeld, J. Math. Phys. {\bf 10}, 1458 (1969).
\bibitem{Hansch1998} T. Esslinger, I. Bloch, and T.W. Hansch, Phys. Rev. A \textbf{58} (4), 2664 (1998).
\bibitem{Foot2002} N.R. Thomas, A.C. Wilson, and C.J. Foot, Phys. Rev. A \textbf{65}, 063406 (2002).
\bibitem{Minguzzi2005} A. Minguzzi and D.M. Gangardt, Phys. Rev. Lett. \textbf{94}, 240404 (2005).
\bibitem{Kagan1996} Y. Kagan, E.L. Surkov, and G.V. Shlyapnikov, Phys. Rev. A \textbf{54}, R1753 (1996).
\bibitem{DGO2009} J.G. Muga, X. Chen, A. Ruschhaupt, and D. Gu\'ery-Odelin, J. Phys. B: At. Mol. Opt. Phys. \textbf{42}, 241001 (2009).
\end{thebibliography}
\end{document}